\documentclass{aa}   
\usepackage{graphics}

\begin{document}

\title{Modelling the ENLR in NGC~1068.} 
\author{L.S.Nazarova,         \inst{1,3}
P.T.O'Brien,                  \inst{1}
M.J.Ward                      \inst{1}
and P.M. Gondhalekar          \inst{2}        
}

\offprints{P.T. O'Brien}

\institute {Department of Physics \& Astronomy, University of Leicester,
University Road, Leicester, LE1 7RH, U.K.
\and
Rutherford Appleton laboratory, Chilton, OXON, OX11 0QX, U.K.
\and
Astronomical Society, Sternberg Astronomical Institute, Universitetskij 
prosp.13, Moscow, 119899, Russia}

\thesaurus{11(11.01.2; 11.17.2; 02.18.7; 11.09.1 NGC~1068)}

\date{Received July 1997 / Accepted October 24 1997}

\authorrunning{Nazarova et al.}
\maketitle 

\begin{abstract}
We present photoionization models of the high excitation gas in the
Extended Narrow Line Region (ENLR) of NGC~1068. The ENLR line fluxes have
been calculated allowing for attenuation of the central-source ionizing
continuum as a function of distance from the centre. Diffuse continuum
emission from low density ENLR gas is included as an important secondary
source of ionization. The observed high excitation emission further than
25 arcsec from the centre of NGC~1068 can be fitted by photoionization
models using a central-source luminosity of $3.6\times10^{44}$ erg
s$^{-1}$ between 10$^{14.6}$ and 10$^{18.4}$ Hz, with the continuum shape
attenuated by nuclear gas with an integrated column density of
N$_{h}=10^{22}$ cm$^{-2}$. The reflected soft X-ray continuum from the
attenuating gas could be responsible for about 10\% of the observed,
resolved circumnuclear soft X-ray continuum extending out to 15 arcsec
from the centre (Wilson et al 1992).
\keywords{galaxies:active --
                galaxies: nuclei --
                galaxies: individual (NGC~1068) --
                galaxies: Seyfert --
                modelling: galaxies
               }
\end{abstract} 

\section{Introduction}

Spectropolarimetric observations of the Seyfert~2 galaxy NGC~1068 show
broad, polarized FeII and Balmer emission lines in the nuclear spectrum
(Antonucci \& Miller 1985; Snijders et al. 1986; Miller \& Goodrich 1990;
Miller et al. 1991; Tran et al. 1992). These results suggest that this
galaxy actually has a Seyfert~1 nucleus which is obscured from direct
view but is seen in light scattered from, and hence polarized by,
obscuring material which may be arranged in the form of a disk or torus
surrounding the central nucleus.

Inspection of the ENLR spectra shows the existence of ionized gas
emitting high-excitation \ion{[Ne}{v}]$\lambda 3425$ and
[\ion{O}{iii}]$\lambda 5007$ lines at an approximate distance of between
15$^{\prime \prime}$ and 50$^{\prime \prime}$ from the centre of NGC~1068
(Evans \& Dopita 1986; Bergeron et al. 1989). Lower ionization lines such
as [\ion{O}{ii}]$\lambda 3727$ and [\ion{N}{ii}]$\lambda 6583$ have
intensity ratios relative to H${\alpha}$) which are similar to those
seen in \ion{H}{ii} regions, and do not reflect the degree of ionization
which might be expected based on the detection of the
[\ion{Ne}{v}]$\lambda 3425$ line (Evans \& Dopita 1986). Furthermore the
spectra of the ENLR show a low intensity of [\ion{O}{iii}]$\lambda 5007$
and \ion{He}{ii}$\lambda 4686$ lines compared to the intensity of
[\ion{Ne}{v}]$\lambda 3425$.

Evans and Dopita (1986) modelled the ENLR spectra in NGC~1068 with two
emission components: \ion{H}{ii} regions and highly-ionized low density
gas. They also suggested that the continuum seen by the high-excitation
gas has a turn-on energy (simulating a photo-electric cut-off due to
absorption by intervening material closer to the centre) varying over the
range 20-60 Ryd. This decreases the intensity of the
\ion{He}{ii}$\lambda4686$ and [\ion{O}{iii}]$\lambda5007$ lines compared
to the intensity of [\ion{Ne}{v}]$\lambda3425$. Bergeron et al. (1989)
used optical emission-line ratios to infer that the ionizing continuum
seen by the ENLR gas may be two orders of magnitude larger than the
observed soft X-ray continuum.

Nazarova (1995) modelled the high-excitation ENLR gas in NGC~1068 with a
central power-law continuum and an extended source of ionization. The
extended source was assumed to be stars of temperature 80\,000K located
1--2~kpc from the centre (or 14--28 arcsec adopting 1 arcsec = 72pc
(Tully 1988)). The central region of NGC~1068 does have a ring of very
luminous \ion{H}{ii} regions 13 arcsec from the centre (Snijders et al. 1982;
Bruhweiler et al. 1991). Adding an additional stellar source changes the
shape of the incident spectrum which ionises the ENLR gas by adding a
bump at 2.5 Ryd (corresponding to the peak emission energy of a star with
temperature 80\,000K). This bump changes the population of the ionization
stages. The O$^{+2}$ moves to the higher stages O$^{+4}$ and O$^{+5}$ and
He$^{+}$ to He$^{++}$. Similarly Ne$^{+4}$ moves to Ne$^{+5}$, although
this change is small due to the higher ionization threshold. Overall,
adding the stellar component leads to stronger emission in
[\ion{Ne}{v}]$\lambda3425$ compared to that of [\ion{O}{iii}]$\lambda5007$ and
\ion{He}{ii}$\lambda4686$.

Although modelling the ENLR spectra of NGC~1068 with an additional
stellar ionization source helps alleviate the ``energy deficit'' problem,
there still exists a problem connected with the geometry of the extended
region. Since there are many galaxies with ENLRs which show strong
emission in the high-excitation [\ion{Ne}{v}]$\lambda 3425$ line (Binette
et al. 1996), it seems unlikely that they all have a strong contribution
from a ring of very luminous \ion{H}{ii} regions at just the right location
around the nucleus.

Recently, Binette et al. (1996) proposed that the high-excitation line
strengths in ENLRs, together with other problems such as low electronic
temperatures and the small range in the \ion{He}{ii}/H$\beta$ ratio,
could be successfully solved by assuming that the ENLR contains a
combination of matter-bounded and ionization-bounded clouds. These two
cloud populations have different spectra and therefore can be combined to
reproduce a wide range of observed ENLR spectra. A principal result was
that the ionization-bounded clouds are photoionized by radiation from the
central-source which has been attenuated by the matter-bounded component.

In this paper we explore the advantages of such multi-component gas
models in attempts to fit the observed ENLR spectrum of NGC~1068,
adopting the shape of the central continuum as given by Pier et al.
(1994). We also examine the effect of different central luminosities and
different amounts of attenuation of these continua on the predicted ENLR
spectrum. Models of the observed ENLR emission are discussed in Section
2. We examine in Section 3 if the model which best fits the ENLR line
emission can also explain the observed extended soft X-ray continuum. The
conclusions are given in Section 4.

\section{Modelling the ENLR spectra.}

\subsection{The ionizing continuum}

The shape of the central-source continuum in NGC~1068 has been derived
from two sources. The UV and X-ray continua ware assumed from Pier et al.
(1994), while the rest of the continuum shape was taken to follow that of
the canonical AGN continuum given in Mathews and Ferland (1987). These
two continua are shown in Fig.1.

\begin{figure}
\resizebox{\hsize}{!}{\includegraphics{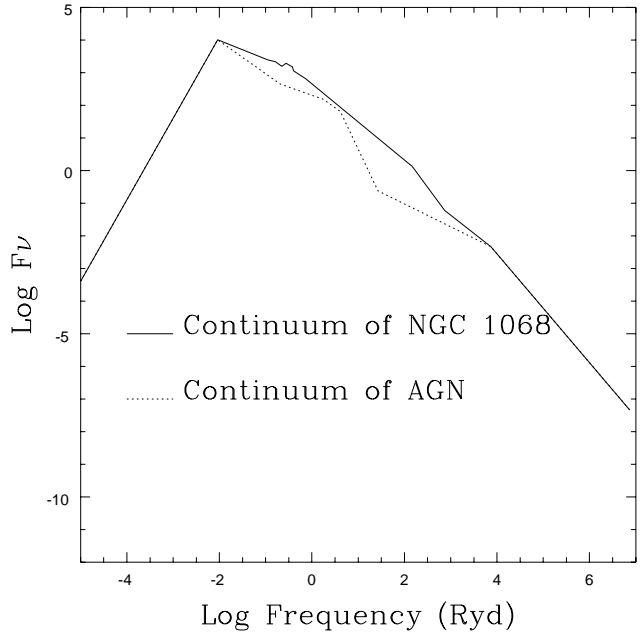}}
\caption{The adopted continuum of NGC~1068 compared to the
canonical AGN continuum of Mathews and Ferland (1987). Both continua are
plotted in units of Photons Ryd$^{-1}$ cm$^{-2}$ s$^{-1}$ with arbitrary
offsets.}
\end{figure}

The bolometric flux from the scattering material in the spectral range
10$^{14.6}$--10$^{18.4}$ Hz is $1.5 \times 10^{-10}$ erg cm$^{-2}$
s$^{-1}$ (Pier et al. 1994), corresponding to a luminosity of
$3.6\times10^{42}$ erg s$^{-1}$ for an adopted distance to NGC~1068 of
14.4~Mpc (Tully 1988). Adopting this as a minimum central-source
luminosity, our photoionization models were calculated within the
luminosity range $3.6\times10^{42}$ to $3.6\times10^{44}$ erg s$^{-1}$
(Log~L~$=42.55$--44.55). The luminosity upper limit was chosen to be
approximately half the IR luminosity of NGC~1068
(L$_{IR}=7.3\times10^{44}$ erg s$^{-1}$) found by Telesco and Harper
(1980), as about half of the IR emission could be due to dust heated by
stars.

\subsection{Location of the [\ion{Ne}{v}] zone}

Evans and Dopita (1986) and Bergeron et al. (1989) detected extended
high-excitation emission from two knots in NGC~1068 located between 29.4
and 43.5 arcsec and between 24.7 and 40.5 arcsec north-east of the
centre. Apart from these, there is also a small [\ion{Ne}{v}] emission
patch in the south-west between 18.2 and 28.8 arcsec from the centre.
However the strong [\ion{O}{iii}] emission extends further out than the
Balmer emission, and is detected up to 45 arcsec from the centre
(Bergeron et al. 1989). In our calculations we have adopted a location
for the high-excitation ENLR region in NGC~1068 (henceforth called the
[\ion{Ne}{v}] zone) between 25 and 45 arcsec from the centre (or between
1.8 and 3.24 kpc).

\subsection{Method}

All calculations were carried out using CLOUDY (version c90.01; Ferland
1991) for a plane-parallel slab of gas with solar abundances. Because the
high-excitation emissivity in the ENLR depends strongly on the shape of
the ionizing continuum, we calculated the line fluxes taking into account
the attenuation of the ionizing continuum by gas at radii between 5 and
25 arcsec from the centre (Section 2.4; Nazarova et al. 1997). The
transmitted portion of the central-source continuum and the diffuse
continuum (emitted by the low density gas) were calculated for 6
different luminosities of the central-source within the luminosity range
given in Section 2.1, with 5 different column densities and different
corresponding gas densities. The sum of the transmitted continuum plus
the diffuse continuum was then taken as the incident continuum at the
starting radius of the [\ion{Ne}{v}] zone (25 arcsec from the centre).
The shape of the unattenuated, central-source continuum and the sum of
the transmitted and diffuse continua are shown in Fig.2. The shape of the
attenuated continua depend on both the luminosity of the central-source
and the column density of the attenuating gas. The attenuation is large
in the EUV/soft X-ray range when the gas is effectively ionization
bounded for the dominant absorption species.

\begin{figure*}
\resizebox{\hsize}{!}{\includegraphics{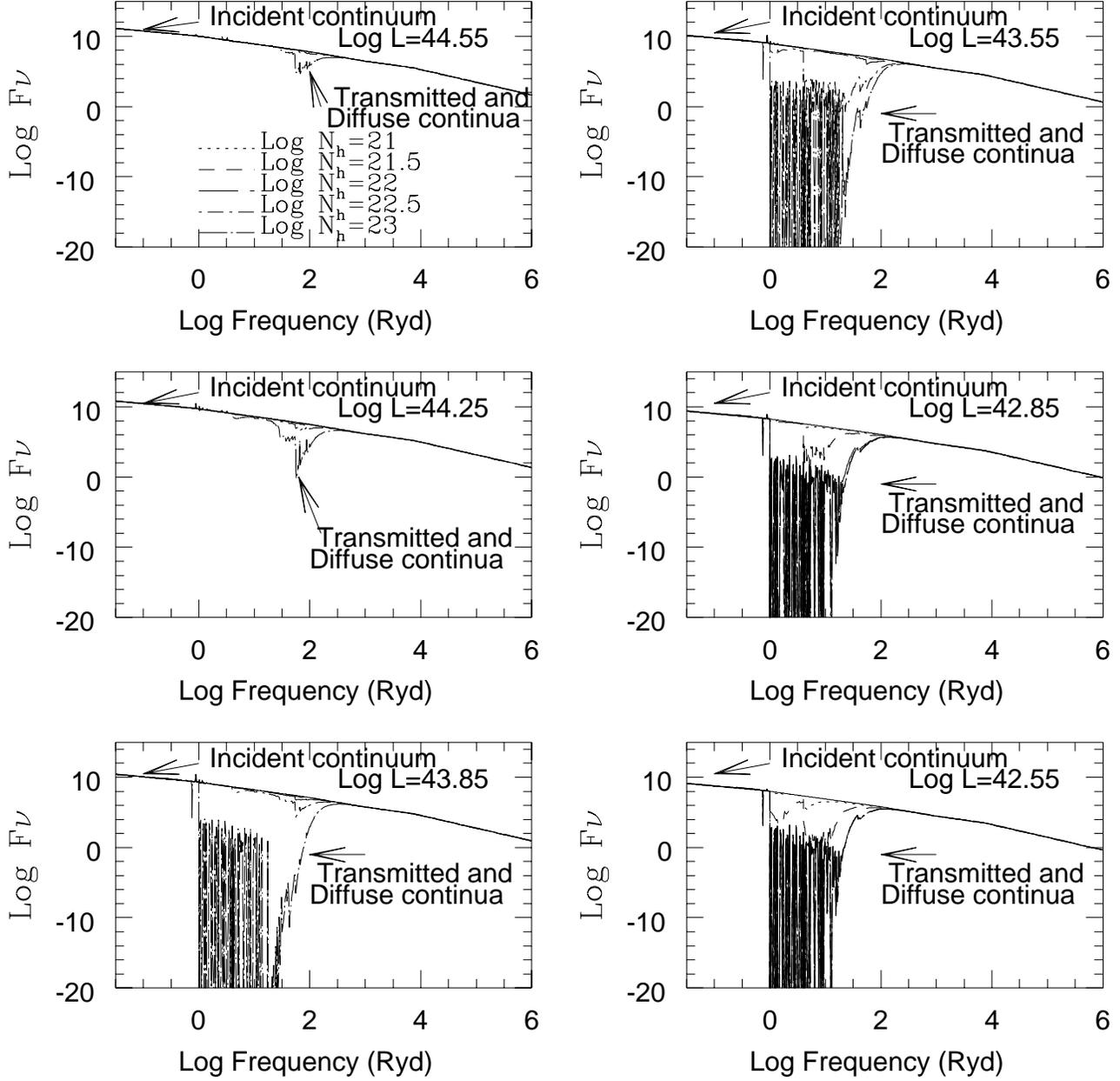}}
\caption{The unattenuated central-source continuum (labelled incident)
together with the sum of the transmitted and diffuse continua for 6
different luminosities of the central-source. Units are Photons
Ryd$^{-1}$ cm$^{-2}$ s$^{-1}$, with an arbitrary constant added to each
continuum for display purposes.}
\end{figure*}

It should be noted that the shape of the sum of the transmitted and
diffuse continua is quite different from the transmitted continuum alone.
Adding the locally-emitted diffuse continua reduces the depth of the gap
in the continuum between approximately 1 and 100 Ryd. The emission of the
high-excitation [\ion{Ne}{v}]$\lambda 3425$ and [\ion{O}{iii}]$\lambda\lambda4959$, 5007
lines depend on the shape of the continuum in this region as it controls
the parent ion populations. The electron temperature of the [\ion{Ne}{v}] zone is
shown in Fig.3 for different luminosities of the central-source. The
sharp drop in temperature after column densities 10$^{22}$ cm$^{-2}$ and
10$^{22.5}$ cm$^{-2}$ respectively for low luminosities
(Log~L~$=42.55$--42.85) is not shown. For high luminosities
(Log~L~$=43.85$--44.55) the temperature remains high enough to produce
strong emission in the high-excitation lines even for high column
densities (N$_{h}=10^{22-23}$ cm$^{-2}$).

\begin{figure}
\resizebox{\hsize}{!}{\includegraphics{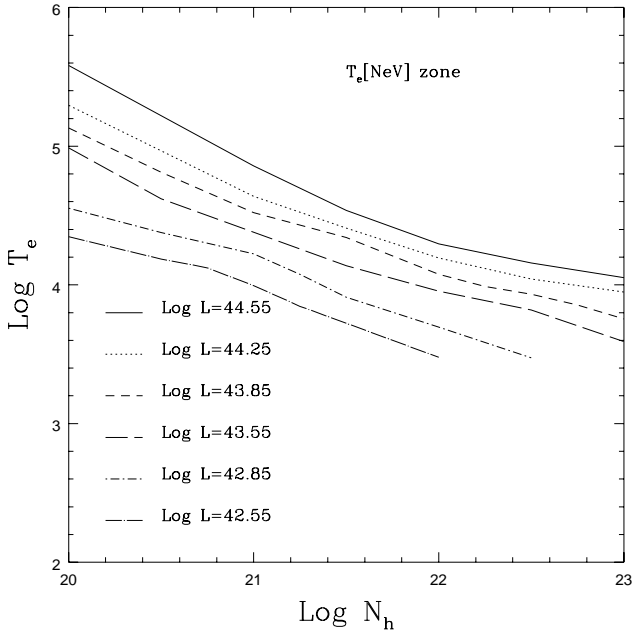}}
\caption{The electron temperature profiles for a homogeneous
envelope of low density gas in the [\ion{Ne}{v}] zone for 
different luminosities of the central-source.}
\end{figure}

\subsection{Attenuating gas geometry}

Since we use in our calculations the sum of the transmitted and diffuse
continua as the incident continuum for the [\ion{Ne}{v}] zone, we need to consider
the geometry of the gas which attenuates the central-source continuum. As
the physical size of the region containing the attenuating gas is fixed
(5--25 arcsec from the centre), varying its total column density N$_{h}$
between 10$^{20}$ and 10$^{24}$ cm$^{-2}$ implies a variation in gas
density between 0.023 and 234 cm$^{-3}$, assuming the gas is homogeneously
distributed and that the filling-factor remains the same.
The starting radius for the region containing the
attenuating gas has been taken to be 5 arcsec, as this is believed to be
the maximum size of the region containing the material obscuring the
central-source from direct view (Pier et al. 1994). The contribution from
scattered central continuum to the illumination of the [\ion{Ne}{v}] zone depends
on the optically thickness for electron scattering ($\tau_{scat}$). The
scattered continuum will also be absorbed in the gas with corresponding
optical thickness ($\tau_{absorb}$). As $\tau_{scat}$ is 10$^{-10}$ times
less than $\tau_{absorb}$ we can neglect the contribution of the
scattered continuum to the shape of the ionizing continuum for the [\ion{Ne}{v}]
zone.

\subsection{Low density gas in the ENLR of NGC~1068}

The high-excitation [\ion{Ne}{v}]$\lambda 3425$ line and part, possibly most, of
the [\ion{O}{iii}]$\lambda 5007$ and HeII$\lambda 4686$ lines beyond 25 arcsec
are produced by highly-ionized, low density gas in the ENLR. The emission
from low density gas in the [\ion{Ne}{v}] zone corresponding to various total
column densities is shown in Fig.4 for different luminosities of the
central-source. The two parallel lines show the observed ratios in the
line intensities for different parts of the ENLR of NGC~1068 derived from
Evans and Dopita (1986) and Bergeron et al. (1989). From Fig.4 it appears
that the observed line fluxes could be fitted with a luminosity of the
central-source higher than Log~L~$=43.85$.

\begin{figure*}
\resizebox{\hsize}{!}{\includegraphics{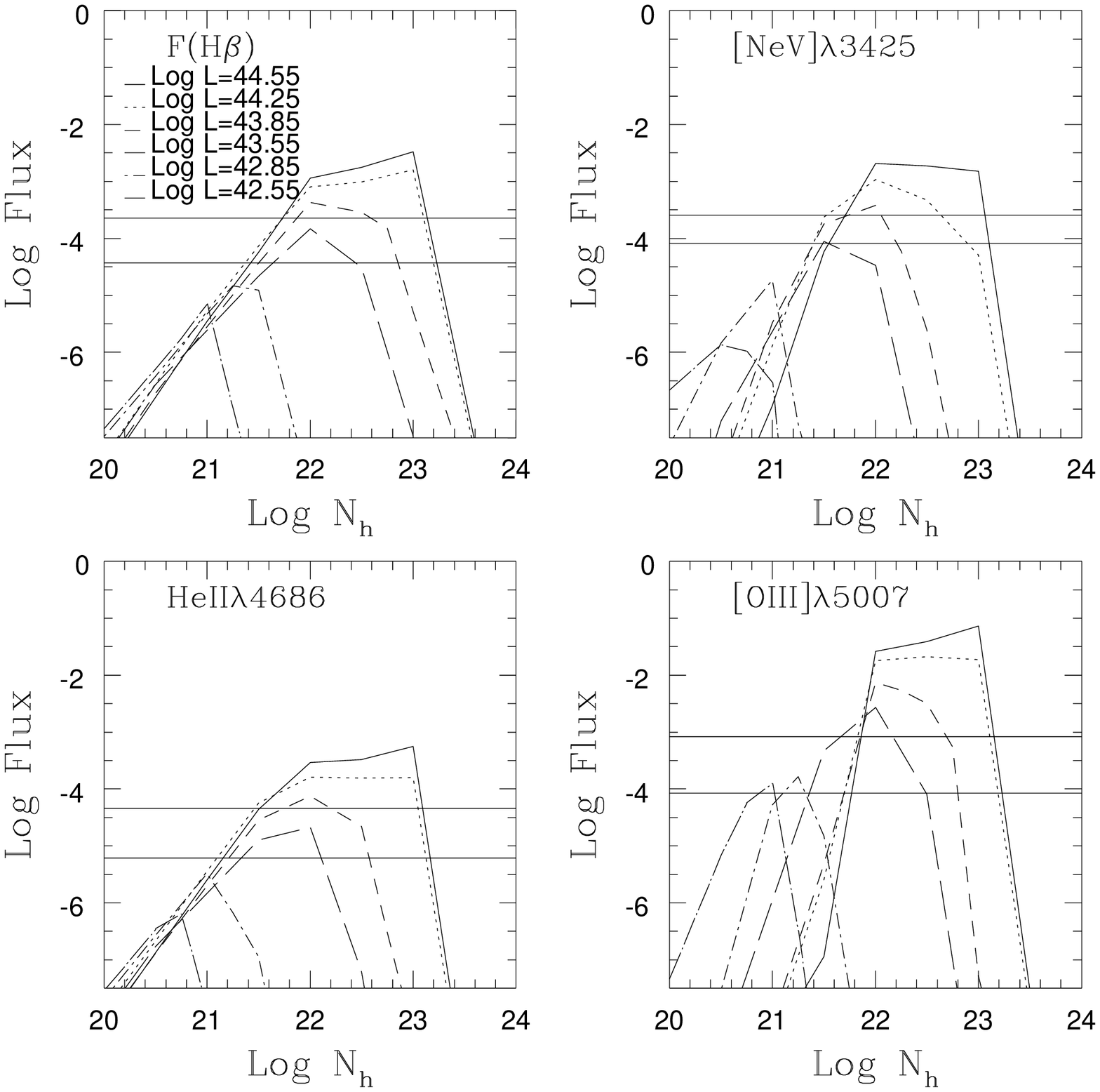}}
\caption{The theoretical fluxes of the
H$\beta$, HeII$\lambda 4686$, [\ion{Ne}{v}]$\lambda 3425$ and
[\ion{O}{iii}]$\lambda 5007$ lines (in units of erg cm$^{-2}$ s$^{-1}$ at
the ENLR) from low density gas in the [\ion{Ne}{v}] zone versus column
density for different luminosities of the central-source.}
\end{figure*}

Adding emission from high density gas, which could explain the observed
low ionization line emission (discussed below), will increase the
emission in the H$\beta$, [\ion{O}{iii}]$\lambda 5007$, and HeII$\lambda
4686$ lines, but makes little contribution to the emission in
[\ion{Ne}{v}]$\lambda 3425$. Therefore, we rule out luminosities below
Log~L~$=43.85$ as they fail to explain the minimum observed flux in
[\ion{Ne}{v}]$\lambda 3425$ even with a filling-factor of 1. For
luminosities higher than Log~L~$=43.85$ the filling-factor in the
[\ion{Ne}{v}] zone might be less than 1 in order to fit the observed line
intensities.

We note that the predicted [\ion{Ne}{v}]$\lambda 3425$ flux is lower than
[\ion{O}{iii}]$\lambda 5007$ for all luminosities, in agreement with the
observed F($\lambda 3425$)/F($\lambda 5007$) line ratio which varies
between 0.26 and 0.58 in different parts of the ENLR (Bergeron et al.
1989). As can be seen in Fig.5, a small change in column density produces
a quite dramatic change in the F($\lambda 3425$)/F($\lambda 5007$) ratio.
This is due to the approximately 1.7 times higher photon energy needed to
produce [\ion{Ne}{v}] than [\ion{O}{iii}]. Hence the former is more
sensitive to the column. The observed F($\lambda 3425$)/F($\lambda 5007$)
ratio can be fitted for a small range of ``critical'' column densities
from 10$^{20.5}$ cm$^{-2}$ to 10$^{22}$ cm$^{-2}$ for the luminosity
range considered here.

\begin{figure}
\resizebox{\hsize}{!}{\includegraphics{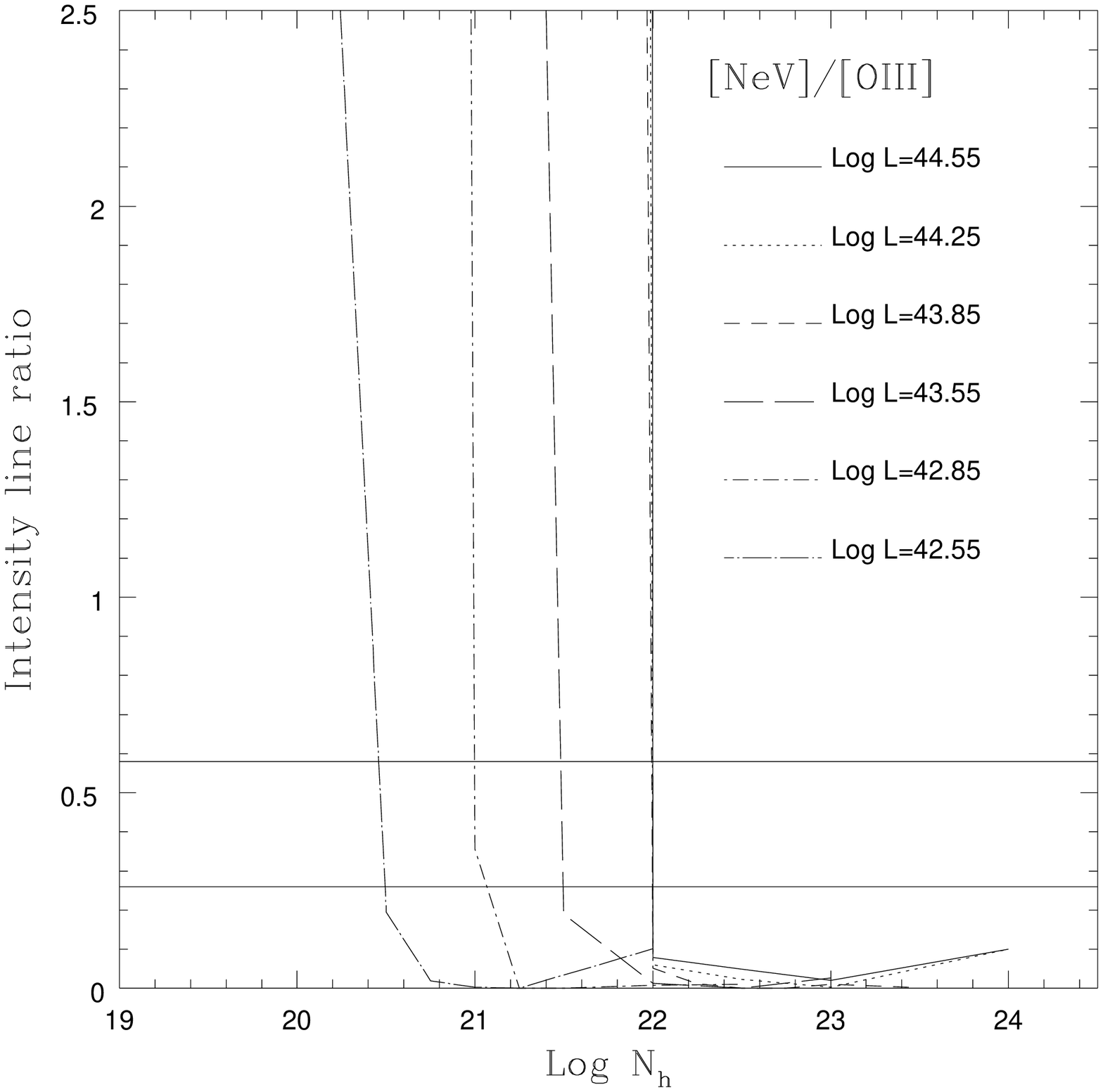}}
\caption{The theoretical and observed
[\ion{Ne}{v}]$\lambda 3425$/[\ion{O}{iii}]$\lambda 5007$ line ratio
versus column density for different luminosities of the central-source.}
\end{figure}

\subsection{Two-component model of the [\ion{Ne}{v}] zone}

The presence of strong emission in low ionization lines such as H$\beta$,
[\ion{O}{ii}]$\lambda 3727$, [\ion{O}{i}]$\lambda 6300$,
[\ion{N}{ii}]$\lambda\lambda 6548$, 6583, and
[\ion{S}{ii}]$\lambda\lambda 6717$, 6731 indicates emission from low
filling-factor high density gas, similar to \ion{H}{ii} regions, in the
ENLR of NGC~1068. This emission is in addition to that from low density
gas. The emission from high density gas was therefore also calculated for
the 6 different luminosity models. Because the most probable column
density obtained from analysis of the low density gas is N$_{h}=10^{22}$
cm$^{-2}$ (Fig.5), the sum of transmitted and diffuse continua for this
column density was used to calculate the continuum incident on the high
density gas in the [\ion{Ne}{v}] zone.

\begin{table*}
\caption[]{The logarithms of the observed surface brightness of
emission lines scaled to the distance of the [\ion{Ne}{v}] zone and the modelled
lines fluxes (in units of erg cm$^{-2}$ s$^{-1}$) for three values of the
luminosity of the central-source.} \label{TabSecInst} \[
\begin{tabular}{clcccc}
\hline
\noalign{\smallskip} 
$\lambda$ &Ion &Observed\ \ &Log~L$=44.55$&Log~L~$=44.25$&Log~L~$=43.85$\\
\noalign{\smallskip}
\hline
       \noalign{\smallskip}

3425&[\ion{Ne}{v}]&$-$3.59 -- $-$4.09&$-$4.18&$-$4.30&$-$4.42\\ 
3727&[\ion{O}{ii}]&$-$3.29 -- $-$4.10&$-$4.10&$-$4.22&$-$4.76\\
3869&[\ion{Ne}{iii}]&$-$3.80 -- $-$4.50&$-$4.28&$-$4.10&$-$4.35\\
4363&[\ion{O}{iii}]&$\leq$$-$4.30 -- $-$5.45\phantom{$\leq$}&$-$4.88&$-$4.97&$-$5.12\\
4686&\ion{He}{ii}&$\leq$$-$4.34 -- $-$5.21\phantom{$\leq$}&$-$4.97&$-$5.00&$-$5.10\\
4861&H$\beta$&$-$3.63 -- $-$4.43&$-$4.11&$-$4.13&$-$4.06\\
4959&[\ion{O}{iii}]&$-$3.77 -- $-$4.34&$-$3.54&$-$3.53&$-$3.59\\
5007&[\ion{O}{iii}]&$-$3.08 -- $-$4.07&$-$3.08&$-$3.08&$-$3.13\\
6300&[\ion{O}{i}]&$\leq$$-$4.60 -- $-$5.30\phantom{$\leq$}&$-$4.11&$-$4.23&$-$3.33\\
6548&[\ion{N}{ii}]&$-$3.95 -- $-$4.82&$-$4.42&$-$4.50&$-$4.73\\
6563&H$\beta$&$-$3.28 -- $-$4.99&$-$3.61&$-$3.70&$-$3.59\\
 6583&[\ion{N}{ii}]&$-$3.45 -- $-$4.17&$-$3.95&$-$3.70&$-$4.25\\
6717&[\ion{S}{ii}]&$-$3.96 -- $-$4.70&$-$4.11&$-$4.22&$-$4.28\\
6731&[\ion{S}{ii}]&$-$4.18 -- $-$5.07&$-$3.94&$-$4.00&$-$4.10\\
            \noalign{\smallskip}
            \hline
HDG/LDG&&&1&1&1\\
Log~N$_{e}$(HDG)&&&4&4&4\\
Log~N$_{e}$(LDG)&&&0.37&0.37&0.37\\
%Log~N$_{h}$&&&&&&\\
%N$_{e}$&&&&&\\
Log~f&&&$-$1.50\phantom{$-$}&$-$1.33\phantom{$-$}&$-$1.00\phantom{$-$}\\ 
\noalign{\smallskip} 
\hline 
        \end{tabular}
      \]     
\begin{list}{}{}
\item[] 
HDG/LDG - The relative contributions from High and Low density gas 
to the H$\beta$ flux;

N$_{e}$(LDG) - The density of Low density gas in units cm$^{-3}$; 

N$_{e}$(HDG) - The density of High density gas in units cm$^{-3}$; 

f - The filling-factor.
\end{list}
\end{table*} 

Composite photoionization models for three luminosities of the
central-source are given in Table~1. The observed and predicted line
fluxes in the [\ion{Ne}{v}] zone from 1 cm$^{2}$ of the surface (the
surface brightness scaled to the distance of the [\ion{Ne}{v}] zone) are
given together with the relative contributions of the high (HDG) and low
(LDG) density gas (HDG/LDG) to the H$\beta$ flux, their densities
(Log~N$_{e}$(LDG) and Log~N$_{e}$(HDG)) and the filling-factor f for the
[\ion{Ne}{v}] zone. The filling-factor was estimated by comparison of the
observed and predicted H$\beta$ flux. Table~1 shows that most appropriate
luminosity for the prediction of the observed line fluxes is
Log~L~$=44.55$.

Binette et al. (1996) note that all of the NLR photoionization models
which fit the strong emission lines predict an [\ion{O}{iii}]$\lambda
4363$/[\ion{O}{iii}]$\lambda 5007$ ratio $< 0.01$, which is smaller than
the observed ratio $\sim$ 0.015. The electron temperature obtained from
the observed [\ion{O}{iii}]$\lambda 4363$/[\ion{O}{iii}]$\lambda 5007$
ratio is often higher than the equilibrium temperature of models
calculated with densities less than 10$^{4}$ cm$^{-3}$. In our models
this line ratio is 0.0157 for a luminosity Log~L~$ = 44.55$, consistent
with the typically observed ratio.

If we suggest a higher luminosity than Log~L~$ = 44.55$ for the
central-source it increases the emission in the high-excitation lines and
simultaneously decreases the emission in [\ion{O}{i}]$\lambda 6300$. However it
seems unreasonable to increase the proposed luminosity of the 
central-source more than the observed IR luminosity of Log~L$_{IR}=44.88$
(Telesco \& Harper 1980). Therefore we believe that the best fit for the
observed ENLR spectra in NGC~1068 corresponds to a central-source
luminosity of Log~L~$= 44.55$.

\section{The reflected continuum}

Investigation of the NGC~1068 nuclear region in soft X-rays (Wilson et
al.,1992) shows 3 X-ray sources; unresolved emission associated with the
Seyfert nucleus (55\% of the total soft X-ray flux); resolved,
circumnuclear emission extending from the centre to a distance of 15
arcsec (23\%); and large-scale emission up to 60 arcsec (22\%), which has
a morphology and spatial scale similar to that of the starburst disk. As
electrons or dust in the ENLR could reflect some of the central-source
continuum into our line of sight (Heckman et al. 1995; Cid Fernandes \&
Terlevich 1995; Tran 1995), it is interesting to see how much of the
circumnuclear soft X-ray continuum could be nuclear emission scattered
from the proposed attenuating gas with column density N$_{h}=10^{22}$
cm$^{-2}$.

We used CLOUDY to calculate the reflected continuum for different
luminosities. We assume the reflected continuum is the continuum emitted
from the illuminated face of the cloud back in the direction of the
central-source (i.e. into 2$\pi$ sr). For a plane-parallel slab the
proportion of the reflected continuum in the direction to the observer
could vary from $\sim 0.5$ (when the ENLR is located in the plane of the
sky) to zero (when we see the nucleus through the ENLR). The optical
images imply that the galaxy disk, and therefore presumably the ENLR gas,
inclination is 40$^{o}$$\pm$3$^{o}$ (Brinks et al. 1997). This implies
that the fraction of the reflected continuum in the direction of the
Earth is $\approx$ 0.4.

The intensity of the reflected continuum depends on both the luminosity
of the central continuum and the column density of the scattering medium.
As the column density grows with distance from the centre we expect that
the contribution of the scattered light should also increase with 
distance.

\begin{figure}
\resizebox{\hsize}{!}{\includegraphics{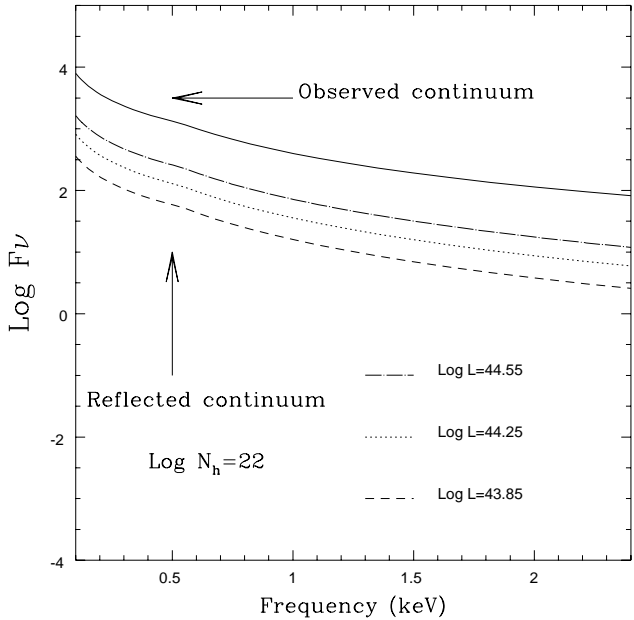}}
\caption{ The observed, resolved continuum and
model reflected soft X-ray continua within 15 arcsec. The continua are
plotted in Photons Ryd$^{-1}$ cm$^{-2}$ s$^{-1}$ on an arbitrary scale.}
\end{figure}

The reflected continua calculated by CLOUDY for the various models are
presented in Fig.6. In each case the reflected continuum comprises the
scattered central-source continuum plus the diffuse continuum emitted by
the gas. If the central-source luminosity in NGC~1068 is Log~L~$=44.55$
then about 10\% of the observed, soft X-ray emission from the
circumnuclear component extending to a distance of about 15 arcsec could
be due to reflected continuum from a column density N$_{h}=10^{22}$
cm$^{-2}$. The majority of the extended soft X-ray continuum in NGC~1068
would be due to a different origin, possibly the 10$^{6}$--10$^{7}$ K gas
seen in the ROSAT HRI soft X-ray image. This gas could result from a hot
outflowing wind driven by the putative central hard X-ray source, or
through shocks driven by the radio jets (Wilson \& Elvis 1997).

\section{Conclusions}

The extended high-excitation emission observed between 25 and 45 arcsec
from the centre in NGC~1068 (the [\ion{Ne}{v}] zone) has been modelled
with a combination of low and high density gas illuminated by a central
ionizing source. Before illuminating the high-excitation gas the
continuum is attenuated by gas located within 25 arcsec. The best fit
model has a central-source luminosity between 10$^{14.6}$ and 10$^{18.4}$
Hz of $3.6\times 10^{44}$ erg s$^{-1}$, a low density gas component with
Log~N$_{e}$ = 0.37 and a high density gas component with Log~N$_{e}$ = 4.
From analysis of the fluxes of the high-exitation emission lines, we find
that the attenuating gas between the centre and the high-excitation
region has an integrated column density of N$_{h}=10^{22}$ cm$^{-2}$. The
reflected soft X-ray continuum from this attenuating gas could explain up
to 10\% of the observed, circumnuclear soft X-ray emission identified by
Wilson et al. (1992) extending from the centre out to a distance of about
15 arcsec.

\begin{acknowledgements}
L.Nazarova wishes to thank the University of Leicester for their
hospitality. The calculations were performed on SUN and DEC workstations
provided by the PPARC Starlink project at Leicester. L.N would also like
to acknowledge support under a NATO grant OUTRG. CRG. 951373
\end{acknowledgements}

\end{document}